\documentclass[aps,pra,twocolumn,superscriptaddress,floatfix,longbibliography]{revtex4}
\usepackage{threeparttable}
\usepackage{amsfonts}
\usepackage{amssymb}
\usepackage{graphicx}
\usepackage{dcolumn}
\usepackage{bm}
\usepackage{amsmath}
\usepackage[colorlinks,linkcolor=magenta,citecolor=blue,urlcolor=blue]{hyperref}
\usepackage{changes}
\begin{document}
	
\preprint{This line only printed with preprint option}

\title{Dynamical properties of quasiparticles in a tunable Kekul\'{e} graphene superlattice}

\author{Xiao-Yu Xiong}
\affiliation{Guangdong Provincial Key Laboratory of Quantum Engineering and Quantum Materials, School of Physics and Telecommunication Engineering, South China Normal University, Guangzhou 510006, China}

\affiliation{Guangdong-Hong Kong Joint Laboratory of Quantum Matter, Frontier Research Institute for Physics, South China Normal University, Guangzhou 510006, China}

\author{Xi-Dan Hu}
\affiliation{Guangdong Provincial Key Laboratory of Quantum Engineering and Quantum Materials, School of Physics and Telecommunication Engineering, South China Normal University, Guangzhou 510006, China}

\affiliation{Guangdong-Hong Kong Joint Laboratory of Quantum Matter, Frontier Research Institute for Physics, South China Normal University, Guangzhou 510006, China}

\author{Qizhong Zhu}
\email{qzzhu@m.scnu.edu.cn}
\affiliation{Guangdong Provincial Key Laboratory of Quantum Engineering and Quantum Materials, School of Physics and Telecommunication Engineering, South China Normal University, Guangzhou 510006, China}

\affiliation{Guangdong-Hong Kong Joint Laboratory of Quantum Matter, Frontier Research Institute for Physics, South China Normal University, Guangzhou 510006, China}

\author{Zhi Li}
\email{lizphys@m.scnu.edu.cn}
\affiliation{Guangdong Provincial Key Laboratory of Quantum Engineering and Quantum Materials, School of Physics and Telecommunication Engineering, South China Normal University, Guangzhou 510006, China}

\affiliation{Guangdong-Hong Kong Joint Laboratory of Quantum Matter, Frontier Research Institute for Physics, South China Normal University, Guangzhou 510006, China}

\date{\today}

\begin{abstract}
We investigate the dynamical properties of quasiparticles in graphene superlattices with three typical Kekul\'{e} distortions (i.e., Kekul\'{e}-O, Kekul\'{e}-Y and Kekul\'{e}-M). On the one hand, we numerically show the visualized evolution process of Kekul\'{e} quasiparticles; while on the other hand, we analytically obtain the centroid trajectory of the quasiparticles, and both of them agree well with each other. The results reveal that the relativistic \textit{Zitterbewegung} (ZB) phenomenon occurs in the Kekul\'{e} systems. Furthermore, through analyzing the frequency of ZB, we unveil the one-to-one relationship between ZB and Kekul\'{e} textures, i.e., the ZB frequenies of Kekul\'{e}-O, Kekul\'{e}-Y and Kekul\'{e}-M quasiparticles feature single, double and six frequencies, respectively. Finally, we propose a scheme to distinguish among different Kekul\'{e} textures from the dynamical perspective. The predictions in this paper are expected to be experimentally verified in the near future, so as to facilitate further research of Kekul\'{e} structures in solid materials or artificial systems. 
\end{abstract}

\maketitle

\section{Introduction}
\label{Sec.1}
Kekul\'{e} (Kek) graphene~\cite{CChamon2000} is a superlattice material formed by periodically manipulating the carbon-carbon (C-C) bond density waves in the hexagonal lattice of carbon atoms. Due to the fact that the primitive cell of Kek superlattice is three times that of a standard graphene, the Brillouin zone of Kek system can fold to become $1/3$ that of standard graphene in reciprocal space. This folding of the Brillouin region will lead to the overlap of the high symmetry points K, K$^{'}$ which formerly featured opposite chirality in graphene, so that quasiparticles with chiral symmetry breaking can be induced~\cite{CHou2007, SRyu2009, XXu2009, CWeeks2010,Kopylov2011,MKharitonov2012, CGutierrez2016, CBao2021}.

Kek-O, Kek-Y and Kek-M are the three types of Kek graphene known so far. First, the C-C bond of Kek-O graphene is of an ``O''-shaped texture in the real space, and the corresponding energy gap possesses a structure of gaped Dirac cone as shown in Fig.~\ref{spectra}(a)~\cite{CChamon2000, CHou2007, VVCheianov2009a, VVCheianov2009b}. Kek-O graphene is well recognized for its topological charge fractionalization phenomenon and other topological properties~\cite{CHou2007, KKGomes2012,LHWu2016,YLiu2017, FLiu2017}. In a recent experiment, C. Bao et al. confirmed by microscopic and spectroscopic measurements in a Li-intercalated graphene that Kek-O distortion can open the energy gap to trigger chiral symmetry breaking~\cite{CBao2021}. Second, as shown in Fig.~\ref{spectra}(b), the Kek-Y structure displays a ``Y''-shaped texture with its linear dispersion being the gapless Dirac cone structure~\cite{OVGamayun2018,JJWang2018, Andrade2019}, which can be experimentally obtained by coating the copper substrate with graphene~\cite{CGutierrez2016,DEom2020}. Finally, as for Kek-M system, the enlargement of the primitive cell has been achieved by periodically adjusting the onsite potential~\cite{JWFVenderbos2016, Herrera2021}. 

On the other hand, \textit{Zitterbewegung} (ZB), as one of the most famous relativistic dynamical effects, has attracted extensive attention in recent years~\cite{ESchrodinger1930,KHuang1952,AOBarut1981,JSchliemann2005,JSchliemann2006, MIKatsnelsona2006,LLamata2007, TMRusin2007,JYVaishnav2008, XDZhang2008, FDreisow2010, GDavid2010, RGerritsma2010, CLQu2013, LJLeBlanc2013, ZLi2015,XDHu2021,ZLi2016,BDora2012,LKShi2013,XShen2022}. Previous studies reveal that ZB is caused by interference between the positive and negative energy states of relativistic Dirac particles. By solving Dirac equation that describes the relativistic free particle, one can find that in addition to the classical initial position term and the velocity term, there is also an oscillation term in the electron displacement expression~\cite{XShen2022}. Therefore, the mean motion trajectory of electron will oscillate along the direction with high frequency and small amplitude. As for a wavepacket with finite width, the trajectory decays rapidly with time, making its experimental observation extremely difficult~\cite{JSchliemann2005,JSchliemann2006,MIKatsnelsona2006,LLamata2007,TMRusin2007,JYVaishnav2008,XDZhang2008,FDreisow2010,GDavid2010,RGerritsma2010,CLQu2013,LJLeBlanc2013,ZLi2015,XDHu2021,ZLi2016,XShen2022}.  

Recent research shows that ZB exists not only in high energy physics but also in condensed matter and artificial systems~\cite{JSchliemann2005, LLamata2007, RGerritsma2010, FDreisow2010,CLQu2013, LJLeBlanc2013}. The past decade have witnessed great theoretical and experimental progress on ZB in the field of quantum simulaton. ZB has been predicted in various condensed matter and artificial systems, including semi-metals~\cite{ZLi2016}, topological insulators~\cite{BDora2012, LKShi2013,XShen2022}, semiconductor nanostructures~\cite{JSchliemann2005, JSchliemann2006, MIKatsnelsona2006}, etc. In 2010, Gerritsma et al. successfully simulated ZB in trapped ions experiment~\cite{RGerritsma2010}, and soon afterwards, Qu et.al. and Le Blanc et al. realized ZB in ultracold atomic lattice systems~\cite{CLQu2013, LJLeBlanc2013}. Thanks to experimental achievements, ZB, once regarded as a mathematical deduction with only the theoretical value, eventually saw the light of practical application. On the other hand, theoretical-experimental schemes such as ZB measurement of exciton dynamics in bilayer graphene~\cite{TMRusin2007}, the genernal dynamical method of measuring the topological invariant in an arbitrary multiband topological systems~\cite{XShen2022} and so on, have been proposed one after another.

The linear dispersion features of the Kek distorted graphene superlattices allow for the dynamical properties of Dirac quasiparticle, which makes possible that ZB occurs in the system. In this paper, we discuss the quasiparticle dynamical properties in three differently-structured Kek graphene superlattices, so as to further explore the properties of Kek system from the dynamical perspective.

The rest of this paper is organized as follows. In Sec.~\ref{Sec.2}, we show the general Hamiltonian of the Kek system. By adjusting relevant parameters, one can obtain Kek-O, Kek-Y and Kek-M texture, respectively. In Sec.~\ref{Sec.3}, we analytically solve, in Heisenberg picture, the expression of the expectation value of the quasiparticle wavepackets' coordinates of Kek-O, Kek-Y and Kek-M systems, and give proof of the existence of ZB. We also conduct the numerical simulation of quasiparticles' evolution, which is in good agreement with the analytical results. Sec.~\ref{Sec.4} is the discussion about the dynamical properties of different Kek phases through Fourier analysis, where the one-to-one relationship between dynamical properties and Kek textures has been established based on the analytical expressions. Conclusion of our work is presented in Sec.~\ref{Sec.5}.

\section{Model}
\label{Sec.2}
\begin{figure}[tbhp]
	\centering	                    
	\includegraphics[width=8cm]{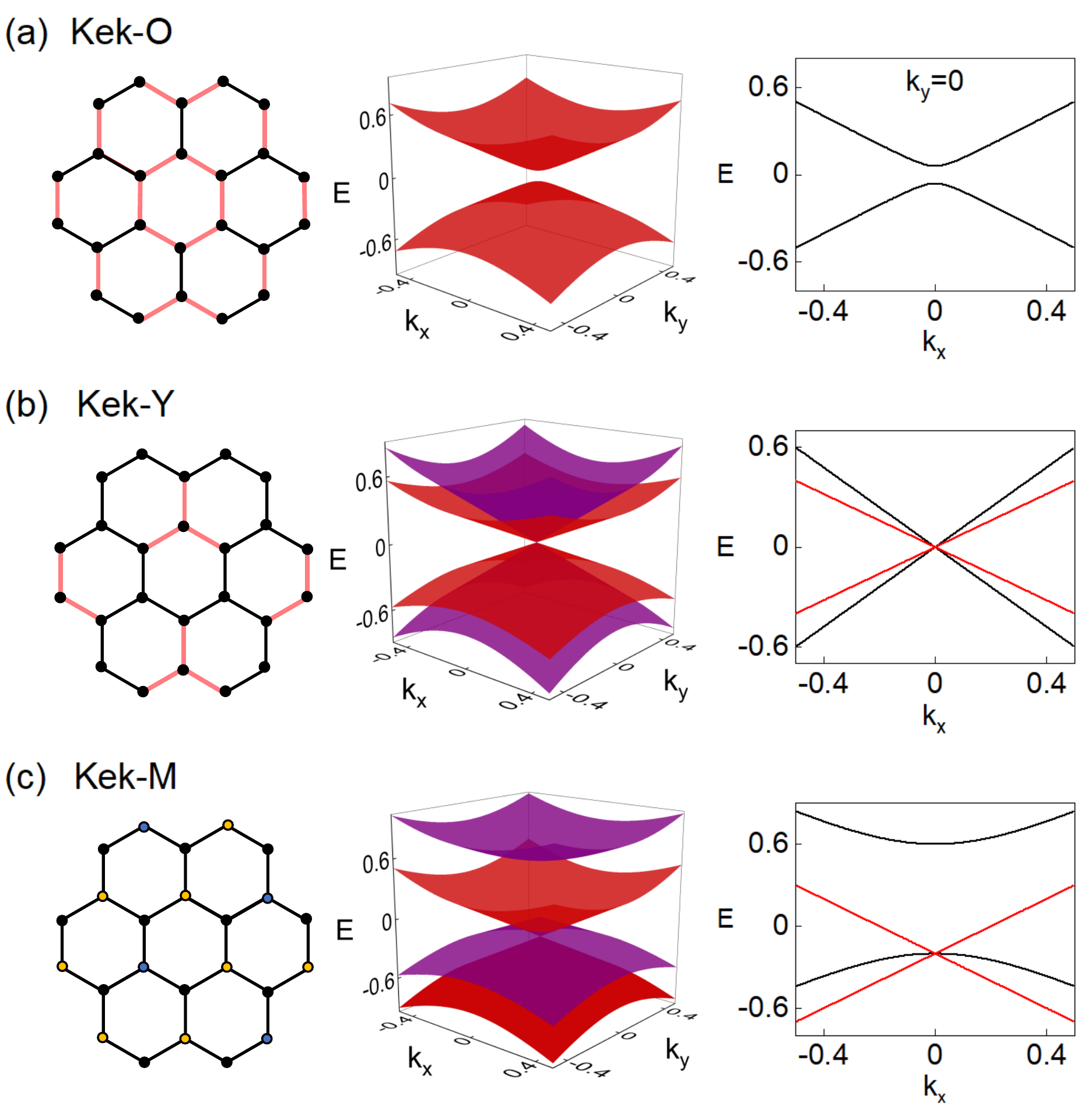}
	\caption{(Color online) (a) (b) and (c) exhibit the lattice structures (the first column) and the corresponding band structures (the other two columns) of Kek-O, Kek-Y and Kek-M system, respectively. Corresponding parameters $v_0=1,~m_0=0.2,~J=1$. For the case of Kek-O and Kek-Y $\Delta=0.2$, while for Kek-M $\Delta=1$. The red and black bonds in the first column indicate different strengths, and the different colored dots represent different on-site energies. }\label{spectra}
\end{figure}

The general low energy effective Hamiltonian of the Kek graphene superlattice reads
%%%
%%%
\begin{equation}\label{Hamil}
H=\left(\begin{array}{cccc}
m_0v_0^2 & v_{0} k_{-} & \eta\Delta Q_{\nu,+}^{*} & 2\Delta m_0v_0^2  \\
v_{0} k_{+} & -m_0v_0^2 & 0 & \eta\Delta Q_{\nu,-}^{*} \\
\eta\Delta Q_{\nu,+} & 0 & -m_0v_0^2 & v_{0} k_{-} \\
2\Delta m_0v_0^2 & \eta\Delta Q_{\nu,-} & v_{0} k_{+} & m_0v_0^2
\end{array}\right),
\end{equation}
%%%
%%%
where $k_{\pm}=k_{x} \pm i k_{y}$. $v_0$ and $m_0$ represent Fermi velocity and effective mass of the quasiparticles in the original graphene system, respectively. $J$ is hopping strength and $\Delta$ is Kek coupling intensity, which can be tuned by the periodic C-C bond in experiments. For Kek-M texture, $\Delta_0=\Delta m_0v_0^2/J$ represents valley coupling intensity. $Q_{\nu, \pm}=v_{0}|\nu|\left(\nu k_{x}-i k_{y}\right) \pm 3 J(1-|\nu|)$. By manipulating the parameters $\eta$, $\nu$ and $m_0$, one can obtain three different Kek textures. In specific, Eq.~\eqref{Hamil} corresponds to the Kek-O (Kek-Y) texture when $\eta=1,m_0=0,\nu=0 (\nu=\pm1)$~\cite{OVGamayun2018}, whereas for $\eta=0$, Eq.~\eqref{Hamil} describes the Kek-M characteristic~\cite{JWFVenderbos2016}. Then, one can obtain the corresponding energy band of the three typical Kek textures as 
%%%
%%%
\begin{equation}
E_{\text{Kek-O}}=\pm \sqrt{9 \Delta^{2} J^{2}+v_{0}^{2}k^{2}},
\end{equation}
%%%
%%%
%~~~
%%%
%%%
\begin{equation}
E_{\text{Kek-Y}}=\alpha v_{0}\left(1+\beta \Delta\right) k,
\end{equation}
%%%
%%%
%~~~
%%%
%%%
\begin{equation}\label{Em}
E_{\text{Kek-M}}=\beta \Delta  m_0v_0^2+\alpha \sqrt{v_{0}^{2} k^{2}+m_{0}^2v_0^4\left(1+\beta \Delta   \right)^{2} },
\end{equation}
%%%
%%%
where $k=\sqrt{k_x^2+k_y^2}$, $\alpha,~\beta=\pm$. The corresponding spectra and the snapshot of $k_y=0$ are plotted in the second and third columns of Fig.~\ref{spectra}. The Kek-O band exhibits one Dirac cone structure with energy gap, and the energy band has double degeneracy [see Fig.~\ref{spectra}(a)]. Kek-Y, however, exhibits two gapless Dirac cone with different slopes [see Fig.~\ref{spectra}(b)]. While for the Kek-M structure, it consists of two Dirac cones that are stacked up and down [Fig.~\ref{spectra}(c)]. Note that, the band structure of Kek-M always has a double degeneracy point at the center of the band (see Appendix~\ref{AP1} for details). In particular, triple degeneracy will occur here when $|\Delta|=1$. Since the dynamical property of quasiparticles is defined by the band structure, one can imagine that different Kek systems are justifiably marked by different dynamical properties. In the following section, we will dive deeper into the dynamics of Kek quasiparticles by applying analytical and numerical methods, respectively.

%%%%%
%%%%%
%%%%%
%%%%%

\section{wavepacket dynamics of Kek quasiparticles}
\label{Sec.3}
%%%%%%
%%%%%%
\begin{figure*}[thbp]
	\centering
	% Requires \usepackage{graphicx}
	\includegraphics[width=16cm]{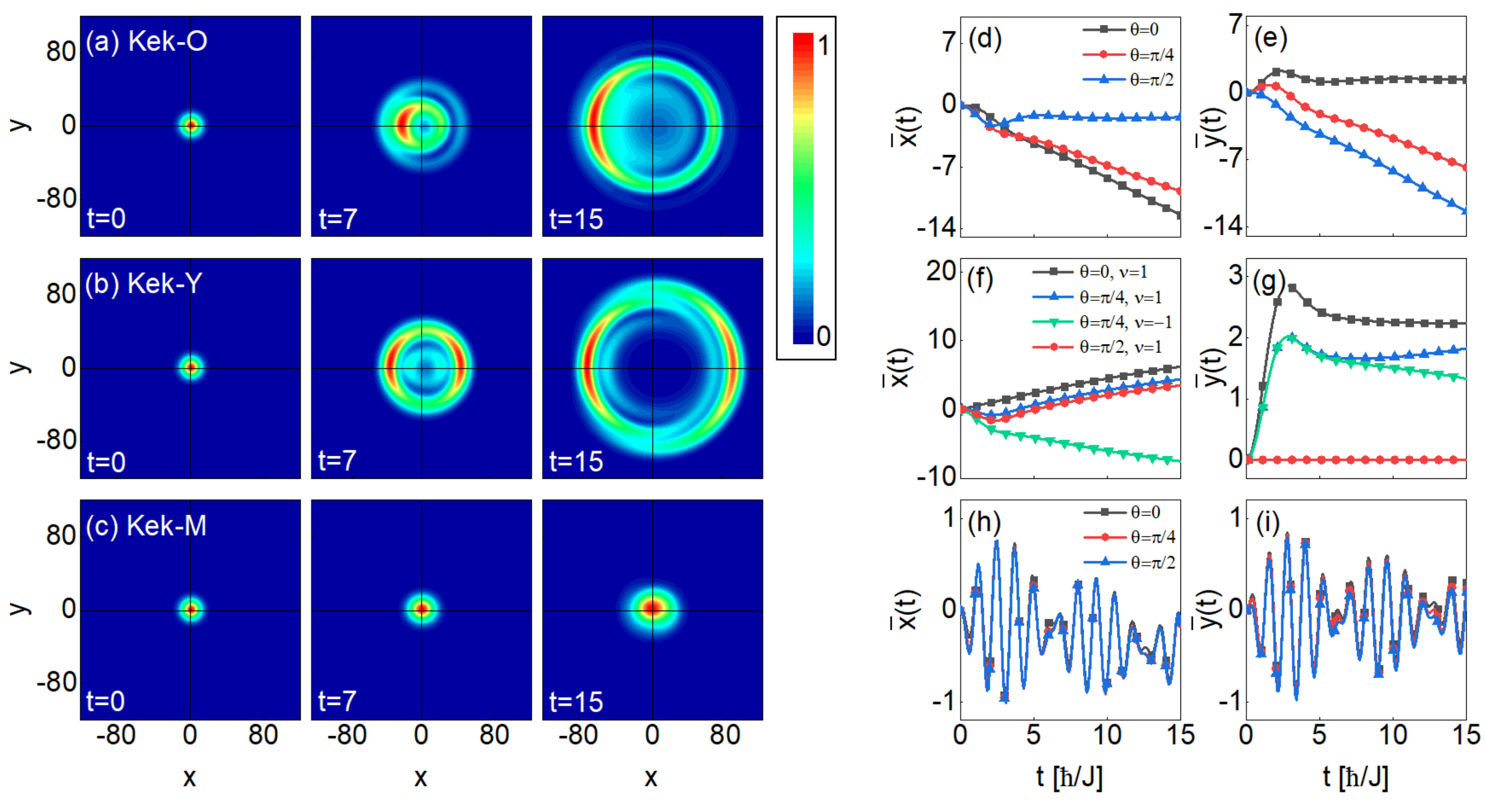}\\
	\caption{(Color online) (a) (b) and (c) Probability distribution of Kek-O, Kek-Y and Kek-M quasiparticles, $|\Psi|^{2}$, at time $t = 0,~7,~15$ with $\theta=0$, $k_0=0.05$, $d=10$. Throughout, $\Delta=0.1,~v_0=5,~J=1$, while $\nu=1$ for Kek-Y and $m_0=0.1$ for Kek-M quasiparticles. (d)-(i) The centroid trajectory of quasiparticles' wavepacket in $x$- and $y$- direction. Rows from top to bottom represent the Kek-O, Kek-Y and Kek-M textures, respectively. The other parameters are marked in the figure.}\label{F2}
\end{figure*}
%%%%%%
%%%%%%
First, in Heisenberg picture, the coordinate operator versus time can be expressed as
%%%
%%%
\begin{equation}\label{eom}
\boldsymbol{\hat{r}}\left(t\right)=U^{\dagger}\boldsymbol{\hat{r}}\left(0\right)U,
\end{equation}
%%%
%%%
where $U=e^{-iHt/\hbar}$ is the evolution operator. Through simple calculation, one can get the time-dependent coordinate operator of Kek Dirac quasiparticles in the following form
%%%
%%%
\begin{equation}\label{coodi}
\begin{aligned}
\boldsymbol{\hat{r}}\left(t\right)=\boldsymbol{\hat{r}_0}+\boldsymbol{\hat{\alpha}}t+\boldsymbol{\hat{\xi}}\left(t\right),
\end{aligned}
\end{equation}
%%%
%%%
where $\boldsymbol{\hat{r}_0}=(x_0, y_0)$ is the initial position term of the quasiparticles' centroid. $\boldsymbol{\hat{\alpha}}=\left(\alpha_{x},\alpha_{y}\right)$ is the corresponding term of drift velocity. One can see that the first two terms of Eq.~\eqref{coodi} correspond to classical motion, whereas $\boldsymbol{\hat{\xi}}=\left(\xi_x,\xi_y\right)$ corresponds to ZB in $x$- and $y$-directions of different Kek cases.

Through analysis, one can obtain the specific expression of the above coefficients for each different Kek texture (see Appendix~\ref{AP2} for details). Moreover, in the Schr\"{o}dinger picture, visualized process of quasiparticles' evolution can be obtained by directly solving the Dirac equation with Hamiltonian Eq.~\eqref{Hamil}. Without loss of generality, a general Gaussian wavepacket is selected as the initial state, which reads
%%%
%%%
\begin{equation}\label{wave}
\Psi=(1 / \sqrt{\pi} d) e^{i k_1 x} e^{i k_2 y}e^{-\left(x^2+y^2\right) / 2 d^2} \Phi,
\end{equation}
%%%
%%%
where $\Phi=(c_1,c_2,c_3,c_4)^T$ is the spinor, and symbol $T$ denotes the matrix transposition. $d$ is the width of the quasiparticle, $k_1=k_0 \cos\theta$ ($k_2=k_0 \sin\theta$) is the wave vector in the $x$- ($y$-) direction, and $\theta$ represents the angle between the wave vector and the $x$-axis. In numerical calculation, we take the spinor as $\Phi=(0,1,0,1)^T$ and set $\hbar=c=1$. Both analytical and numerical results are plotted in Fig.~\ref{F2}.

As shown in Fig.~\ref{F2}(a), the Kek-O quasiparticle exhibits oscillation behavior at the early stage of evolution followed by rapid wavepacket expansion, which is the evidence of ZB. As shown in Fig.~\ref{F2}(b), during the evolution of Kek-Y quasiparticle, similar oscillation behavior can also be found at the early stage of evolution, and then two crescent-shaped structures are formed as quasiparticles expand. This means that the quasiparticles possess two different group velocities as a result of two different slopes existing in the energy band, therefore, parts of the wavepacket spread fast while other parts move slowly.  Besides, as shown in Fig.~\ref{F2}(c), the quasiparticles’ expansion velocity corresponding to Kek-M quasiparticle is slower than that of the other two Kek systems, and the wavepacket, being locked at where it is, just trembles.

Fig.~\ref{F2}(d)-(i) shows the expectation values of $x$ and $y$ coordinates for three different Kek quasiparticles, respectively, where the lines (symbols) denote the analytical (numerical) results.

\begin{figure}[bhtp]
	\centering
	\includegraphics[width=8cm]{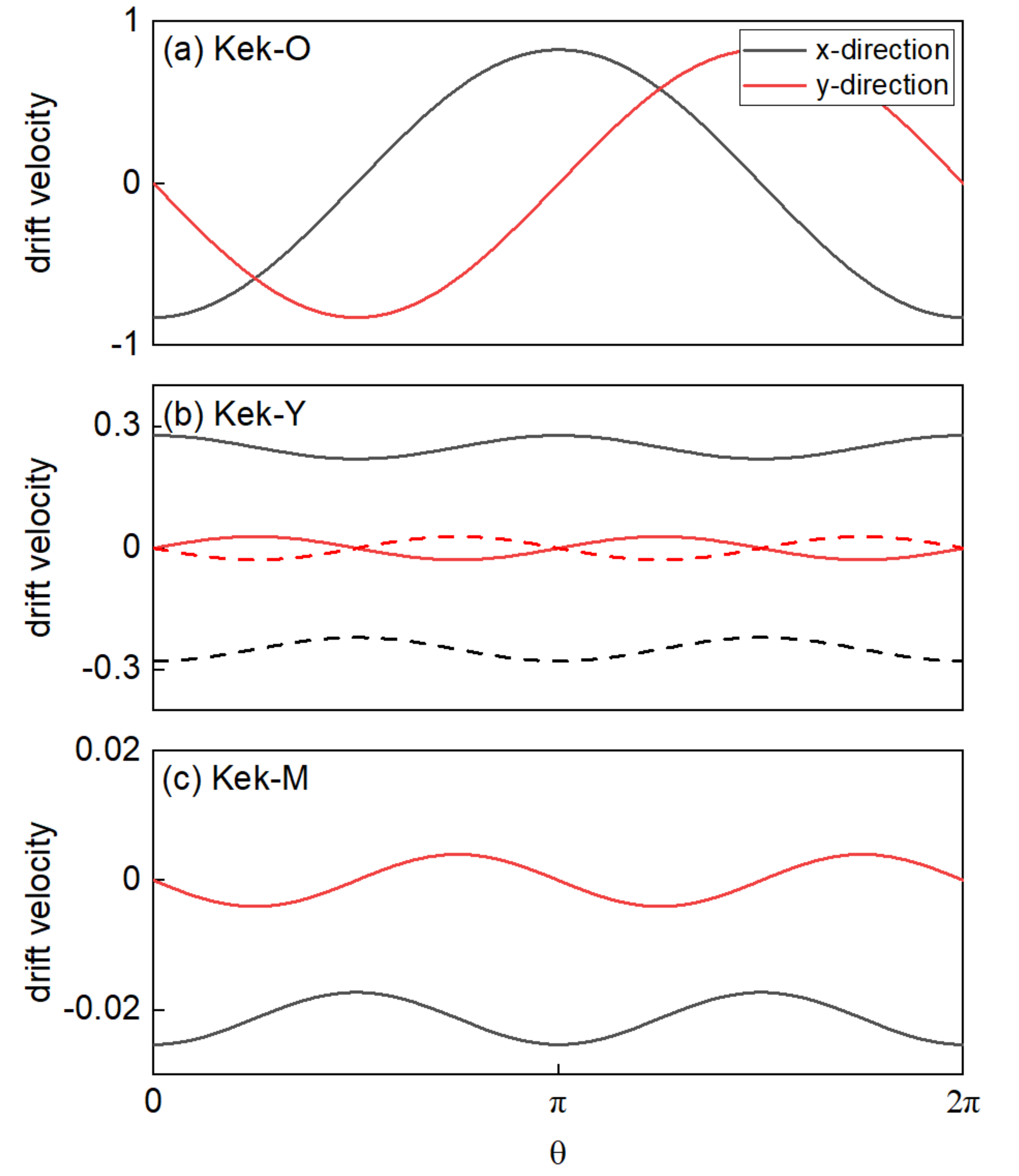}
	\caption{(Color online) The drift velocity versus $\theta$ in the $x$ (black) and $y$ (red) direction for the case of Kek-O (a), Kek-Y (b) and Kek-M (c). The parameters are set the same as Fig.~\ref{F2}. Solid (dashed) line in Kek-Y case [see (b)] corresponds to $\nu=1$ ($\nu=-1$).}\label{drift}
\end{figure}

\textit{The drift velocity.}---With a closer look at Fig.~\ref{F2}(d-i), it is not difficult to find that under the same initial state, different Kek quasiparticles have different drift speeds during the evolution process. How quasiparticls’ drift velocity changes with $\theta$ is plotted in Fig.~\ref{drift}, and the results reveal that periodical change of the drift velocity can be found in all the three typical Kek quasiparticles. To be exact, for Kek-O (Kek-Y and Kek-M), the period of change of quasiparticles’ drift velocity is $2\pi$ ($\pi$). Meanwhile, each type of Kek features a different degree of change of the drift velocity versus $\theta$, i.e., the largest is for Kek-O, smallest for Kek-M, and Kek-Y is in between. Note that, in the case of Kek-Y, the drift of quasiparticles with $\nu=1$ and $\nu=-1$ are in opposite directions, and the velocity in $x$-direction is much greater than that in $y$-direction [see Fig.~\ref{drift}(b)]. In the case of Kek-M, however, since the drift velocity is so small that its effect on ZB is negligibly weak, its effect can be almost just ignored in the evolution process of quasiparticle dynamics [see Fig.~\ref{drift}(c)]. The drift velocity results agree well with the analytic expressions [see Appendix \ref{AP2} for details].

\textit{The frequency.}---Firstly, through theoretical analysis of the analytic results, one can find that ZB of Kek-O quasiparticle features only one frequency, which agrees well with the analysis of corresponding band structure in Fig.~\ref{spectra}. Secondly, one can find that there are two different oscillation frequencies during the dynamical evolution of Kek-Y quasiparticles, one of which is $\Delta$ times of the other one (see Appendix~\ref{AP2} for details). The analytical and numerical results agree very well with each other, which confirms the process of wavepacket evolution. Thirdly, for Kek-M quasiparticles, multifrequency oscillations occur simultaneously in the $x$- and $y$-directions [see Fig.~\ref{F2}(h)(i)]. 

In the next section, we will discuss in detail the corresponding dynamical phenomena of the three Kek quasiparticles by means of Fourier analysis.

\section{Fourier analysis of ZB frequency}
\label{Sec.4}
\begin{figure}[bhtp]
	\centering
	\includegraphics[width=8cm]{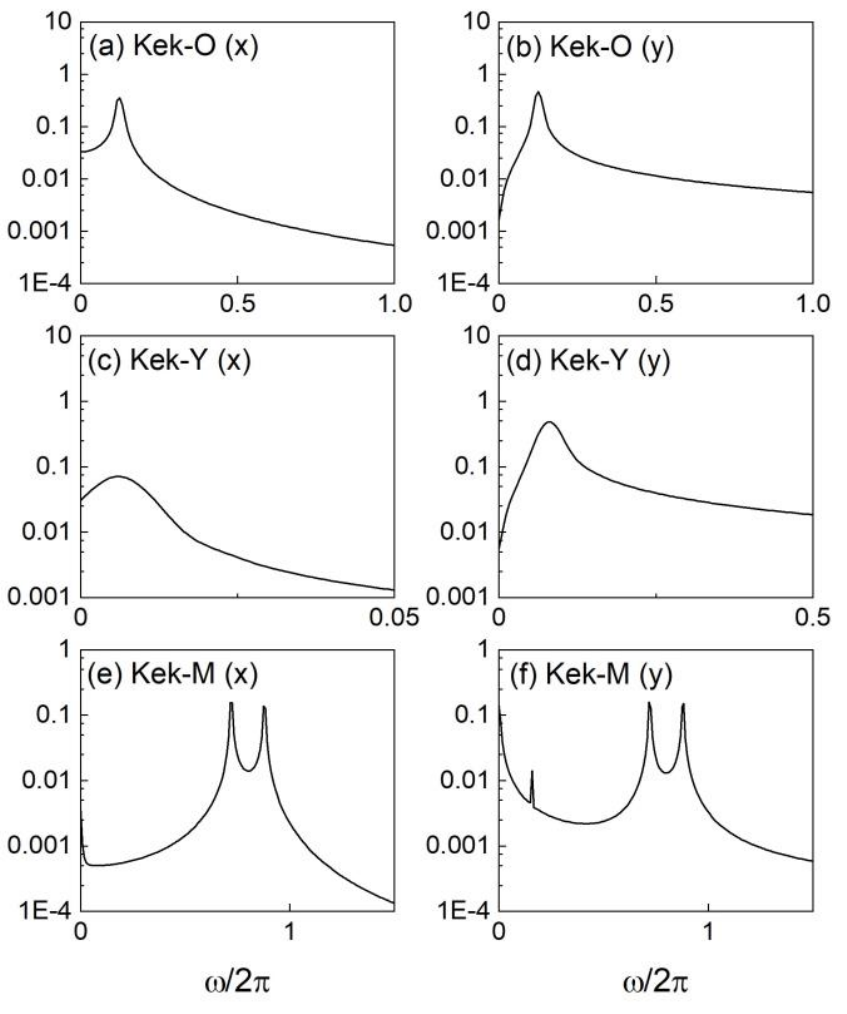}
	\caption{Fourier analysis of the curve of wavepackets centroid with time $t = 180$. Throughout, $\theta=0$, $k_0=0.05$, $\Delta=0.1$, $v_0=5$, $J=1$. To better analyze the frequencies, we set $d=100$. For Kek-Y textures [(c) and (d)], $\nu=1$. For Kek-M texture [(e) and (f)], $m_0=0.1$.}\label{omega}
\end{figure}

To better grasp the characteristics of different types of Kek quasiparticles, we extract the frequency information of quasiparticles' ZB oscillation in the evolution by Fourier analysis. The results are plotted in Fig.~\ref{omega}. 

%%%%%
%%%%%
%%%%%
%%%%%

As shown in Fig.~\ref{omega}(a)(b), two peaks with the same position can be seen in both $x$- and $y$-directions, which indicates the only one frequency of Kek-O ZB. Through analytical calculation of Eq.~\eqref{kekoxi}, the oscillation frequency ($f=\frac{\omega}{2\pi}$) of Kek-O case is about $0.12$, which is consistent with the numerical results of Fourier analysis. Fig.~\ref{omega}(c)(d) show that ZB of Kek-Y quasiparticle features two different peaks in the $x$- and $y$-directions, respectively. On closer inspection, one can find that the frequency in $x$-direction ($f_2=\frac{\omega_2}{2\pi}\approx0.008$) is exactly $\Delta$ ($=0.1$) times of that in $y$-direction ($f_1=\frac{\omega_1}{2\pi}\approx0.08$), which is in consistence with the theoretical prediction [see Appendix~\ref{AP2}]. 

For the case of Kek-M, as shown in Fig.~\ref{omega}(e)(f), one can see three (four) frequency peaks in the $x$($y$)-direction, and there are two non-zero frequencies that coincide in the $x$- and $y$-directions. The lower frequency $f_{31}\approx f_{32}\approx0.7$ represent the differences between $E_1$ ($E_2$) and $E_3$, whereas the higher one $f_{41}\approx f_{42}\approx0.9$ represent the differences between $E_1$ ($E_2$) and $E_4$ [see Appendix~\ref{AP2} for details]. The results reconfirm the correctness of the theory. Since the high symmetry points of the two lower bands $E_1$ and $E_2$ degenerate, the energy differences are almost the same at low energy, resulting in a peak that appears at zero frequency, i.e., $f_{21}\approx 0$. The last frequency $f_{43}\approx0.16$ in $y$-direction stems from level difference between the two upper bands $E_3$ and $E_4$. Although $f_{43}\approx 0.16$ also exists in $x$-direction, since in k space, the value of $\beta_{43}^{x}$ corresponding to the centroid of wavepacket is far smaller than that of $\beta_{43}^{y}$, there will appear a peak in the $y$-direction while a soft one in the $x$-direction [see Appendix~\ref{AP2} details]. Since the two lower bands degenerate at high symmetry point, Kek-M ZB is composed of four frequencies. That is to say, although theoretically there are six frequencies of Kek-M ZB ($\omega_{43},~\omega_{42},~\omega_{41},~\omega_{32},~\omega_{31},~\omega_{21}$), only four of them are detectable in experiments. 

For clarity, We summarize the dynamical properties of different Kek quasiparticles in the following table.
%%%%%%%%%
%%%%%%%%%

\begin{table}[htbp]
	\centering
	\caption{The properties of Kek quasiparticles}
	\label{tab:1}  
	\setlength{\tabcolsep}{5mm}{
	\begin{threeparttable}	
  \begin{tabular}{c | ccc}
			\hline\noalign{\smallskip}	
			System & Kek-O & Kek-Y & Kek-M \\
			\noalign{\smallskip}\hline\noalign{\smallskip}
			$\eta$ & 1 & 1 &0  \\
			$m_0$ & 0 & 0 & Const.  \\
			$\nu$ & 0 & $\pm1$ & -  \\

    \noalign{\smallskip}\hline\noalign{\smallskip}
			$\omega$ & 1 &2 & 6(4 valid)  \\
			\noalign{\smallskip}\hline
	\end{tabular}
\end{threeparttable}}
\end{table}
%%%%%%%%%
%%%%%%%%%

\section{CONCLUSIONS}
\label{Sec.5}
In summary, dynamical properties of Kek-O, Kek-Y and Kek-M quasiparticles are investigated in this paper. The results show that ZB phenomenon exists in Kek quasiparticles due to the Dirac cone structure of the energy band. On the one hand, through numerical simulation, we visualize the evolution of Kek quasiparticle wavepacket with time. On the other hand, through analytical derivation, we obtain the analytical expression of the changing trajectories of the quasiparticle centroid with time. Both of them show in concert the ZB dynamical properties of the system. Further, by Fourier analysis, we obtain the characteristics of ZB frequencies in different Kek systems. In concrete terms, ZB frequencies of Kek-O, Kek-Y and Kek-M quasiparticles are composed of one, two and six frequencies, respectively. Note that, for Kek-M, only four different frequencies can be detected experimentally due to the band degeneracy in the system. Experimentally, based on the one-to-one relationship between frequency and each Kek system, one can determine the Kek texture from the dynamical perspective. Our work will contribute to the deeper understanding of relativistic dynamics, and also bring benefits to the research of Kek distortion.

\begin{acknowledgments}
This work was supported
by the National Key Research and Development Program of China (Grant No. 2022YFA1405300), the National Natural Science Foundation of China (Grant No. 12074180 and No. 12004118), the Innovation Program for Quantum Science and Technology (Grant No. 2021ZD0301705), and the Guangdong Basic and Applied Basic Research Foundation (Grants No. 2021A1515012350, No. 2020A1515110228 and No. 2021A1515010212).
\end{acknowledgments}

\appendix
\section{Band structure of the Kek-M system}\label{AP1}
\begin{figure*}[bhtp]
	\centering
	\includegraphics[width=16cm]{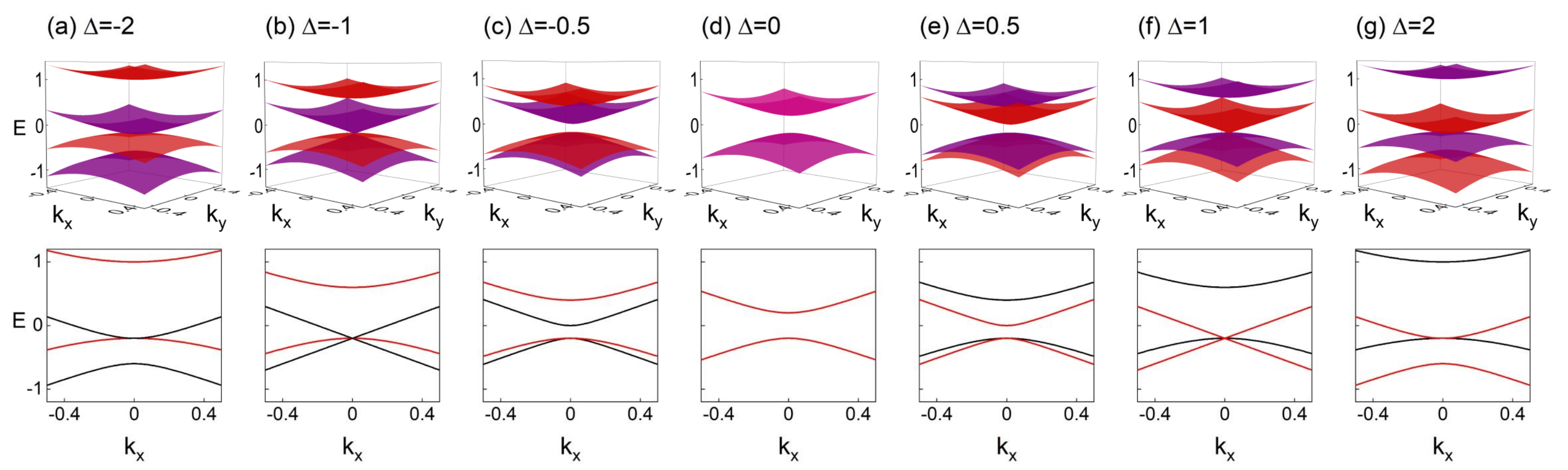}
	\caption{(Color online) Band structures versus $\Delta$. The lower row represent the snapshots of $k_y=0$. The other parameters are fixed, i.e., $m_0=0.2$, $v_0=1$.} \label{fig4}
\end{figure*}

The expression of Kek-M quasiparticles' energy-momentum relationship reads
%%%
%%%
\begin{equation}\label{Em}
E_{\text{Kek-M}}=\beta \Delta  m_0v_0^2+\alpha \sqrt{v_{0}^{2} k^{2}+m_{0}^2v_0^4\left(1+\beta \Delta   \right)^{2} }.
\end{equation}
%%%
%%%
Since $\alpha=\pm$ and $\beta=\pm$, one can get different band structures, which are plotted in Fig.~\ref{fig4}. 

%%%%%%
%%%%%%
%%%%%%
%%%%%%

As shown in the figure, degeneracy always occurs in the energy bands. In most cases, it is double degeneracy, but triple degeneracy can also occur under certain parameters. Without loss of generality, we make a simple analysis of the ZB frequency by the properties at the high symmetry point ($k=0$). Then, one can obtain
%%%
%%%
\begin{equation}\label{En}
	\begin{aligned}
		E_1&=-\Delta m_0v_0^2-m_0v_0^2|1-\Delta|,\\
		E_2&=\Delta m_0v_0^2-m_0v_0^2|1+\Delta|,\\
	    E_3&=-\Delta m_0v_0^2+m_0v_0^2|1-\Delta|,\\
		E_4&=\Delta m_0v_0^2+m_0v_0^2|1+\Delta|,		
	\end{aligned}
\end{equation}
%%%
%%%
Here, we consider the case of $m_{0}>0$ and $\Delta>0$. To get rid of the absolute value sign, let's discuss different cases of $\Delta$.

When $\Delta<1$, Eqs.~\eqref{En} become
\begin{equation}\label{En1}
	\begin{aligned}
		E_1&=-m_0v_0^2,\\
		E_2&=-m_0v_0^2,\\
		E_3&=\left(1-2\Delta\right)m_0v_0^2,\\
		E_4&=\left(2\Delta+1\right) m_0v_0^2,
	\end{aligned}
\end{equation}
which shows a double degeneracy at the center, i.e., $E_1=E_2$ at $k=0$ [see Fig.~\ref{fig4}(e)].

When $\Delta=1$, Eqs.~\eqref{En} become
\begin{equation}\label{En1}
	\begin{aligned}
		E_1&=-m_0v_0^2,\\
		E_2&=-m_0v_0^2,\\
		E_3&=-m_0v_0^2,\\
		E_4&=\left(2\Delta+1\right) m_0v_0^2,
	\end{aligned}
\end{equation}
which shows a triple degeneracy, i.e., $E_2=E_3=E_4$ at $k=0$ [see Fig.~\ref{fig4}(f)].

When $\Delta>1$, Eqs.~\eqref{En} become
\begin{equation}\label{En1}
	\begin{aligned}
		E_1&=\left(1-2\Delta\right) m_0v_0^2,\\
		E_2&=-m_0v_0^2,\\
		E_3&=-m_0v_0^2,\\
		E_4&=\left(2\Delta+1\right) m_0v_0^2,
	\end{aligned}
\end{equation}
which exhibits a double degeneracy of the middle two bands, i.e., $E_2=E_3$ in $k=0$ [see Fig.~\ref{fig4}(g)]. 

One can see that no matter what value $\Delta$ takes, there are always at least two degeneracy points in the band structure of Kek-M. Besides, since ZB frequency is determined by the difference between energy bands, the number of the detectable ZB frequencies in real experiments is always less than the number of interference that occur among different bands.

\section{The derivation details of analytic expressions}\label{AP2}
Firstly, by substituting the Kek-O and Kek-Y Hamiltonian in Eq.~\eqref{eom}, one can get the corresponding expression of coordinates versus time. As for the case of Kek-O, the specific expression can be written as
%%%
%%%
\begin{equation}\label{kekox}
	\begin{aligned}	
		\alpha_{x}=&\frac{v_{0}^{2}}{E^{2}}\left(I_2\otimes \sigma_{x} v_{0}  k_{x}^2+I_2\otimes \sigma_{y} v_{0} k_{y}k_{x}+3 \sigma_{x} \otimes \sigma_{z} \Delta J k_{x}\right) , \\
		\xi_x=&\frac{v_{0} \hbar}{2 E^{2}}\left(-I_2\otimes \sigma_{z} v_{0} k_{y}+
		\sigma_x \otimes \sigma_y 3\Delta J\right)\left[\cos \left(\omega t\right)-1\right]\\
		&+\frac{v_{0} \hbar}{2 E^{3}}\left[-I_2\otimes \sigma_{y}v_0^2  k_x k_y +I_2\otimes \sigma_{x}v_0^2   k_y^2\right.\\ 
		& \left.-\sigma_x\otimes\sigma_z 3 \Delta J v_0 k_x + I_2\otimes \sigma_{x}(3\Delta J)^2\right]
		\sin\left(\omega t\right),
	\end{aligned}
\end{equation}
%%%
%%%
for $x$-direction,
%%%
%%%
\begin{equation}\label{kekoy}
	\begin{aligned}	
		\alpha_{y}=&\frac{v_{0}^{2}}{E^{2}}\left(I_2\otimes\sigma_{x} v_{0} k_{x} k_{y}+I_2\otimes \sigma_{y} v_{0} k_{y}^{2}+3 \sigma_{x} \otimes \sigma_{z} \Delta J k_{y}\right) ,\\
		\xi_y=&\frac{v_{0} \hbar}{2 E^{2}}\left(I_2\otimes \sigma_{z} v_{0} k_{x}-
		\sigma_x \otimes \sigma_x 3\Delta J\right)\left[\cos \left(\omega t \right)-1\right]\\
		&+\frac{v_{0} \hbar}{2 E^{3}}\left[I_2\otimes \sigma_{y}v_0^2  k_x^2 -I_2\otimes \sigma_{x}v_0^2  k_x k_y \right.\\
		&\left.-\sigma_x\otimes\sigma_z 3 \Delta J v_0 k_y + I_2\otimes \sigma_{y}(3\Delta J)^2\right]\sin\left(\omega t\right),
	\end{aligned}
\end{equation}
%%%
%%%
for $y$-direction, where $I_2$ represents a $2\times2$ unit matrix, $\boldsymbol{\sigma}=(\sigma_{x},~\sigma_{y},~\sigma_{z})$ is pauli matrix, $E=|E_{Kek-O}|,~\omega=2E/\hbar$. For the case of the spinor $\Phi=(0,1,0,1)^T$, one can get the expression of $\bar{\xi}_x$ and $\bar{\xi}_y$ as
\begin{equation}
	\begin{aligned}\label{kekoxi}
		\overline{\xi}_x(t)=&\left\langle\Psi_{k}\left|\xi_{x}(t)\right| \Psi_{k}\right\rangle\\
		=&\iint\left\{\frac{\hbar v_0^2k_y}{E^2}\left[\cos(\omega t)-1\right]+\frac{3\hbar Jv_0^2 k_x}{E^3}\sin(\omega t)\right\}\\
		&\times\frac{d^2}{\pi} e^{- d^{2}\left[\left(k_{x}-k_{1}\right)^{2}+\left(k_{y}-k_{2}\right)^{2}\right]}dk_xdk_y,\\
		\overline{\xi}_y(t)=&\left\langle\Psi_{k}\left|\xi_{y}(t)\right| \Psi_{k}\right\rangle\\
		=&\iint\left\{\frac{\hbar v_0^2k_x}{E^2}\left[1-\cos(\omega 	t)\right]+\frac{3\hbar Jv_0^2 k_y}{E^3}\sin(\omega t)\right\}\\
		&\times \frac{d^2}{\pi}e^{- d^{2}\left[\left(k_{x}-k_{1}\right)^{2}+\left(k_{y}-k_{2}\right)^{2}\right]} dk_xdk_y.\\
	\end{aligned}
\end{equation}

It can be seen clearly that there is only one oscillation frequency in the system.

For the case of Kek-Y, the expression reads
%%%
%%%
\begin{equation}\label{kekyx}
	\begin{aligned}
		\alpha_{x}=& \frac{{k}_{x}v_0}{k^2}\left(I_2\otimes \sigma_{x}  {k}_{x}+I_2\otimes \sigma_{y}   {k}_{y}+ \sigma_{x}\otimes I_2v\Delta k_{x}\right.\\
		&\left.+\sigma_{y}\otimes I_2 v\Delta {k}_{y}\right), \\
		\xi_x=&\frac{\hbar k_y}{2k^2}\left\{-I_2\otimes \sigma_z\left[\cos(\omega_1t)-1\right]-\sigma_z \otimes I_2\left[\cos(\omega_2t)-1\right]\right\}\\
		&+\frac{\hbar k_y}{2k^3}\left\{\left[ I_2\otimes \sigma_{x}k_y-I_2\otimes \sigma_{y} k_x\right]\sin\left(\omega_1 t\right)\right.\\		
		&\left.+\left[ \sigma_x\otimes I_2 k_y-\sigma_y\otimes I_2k_x\right]\sin\left(\omega_2 t\right)\right\} ,
	\end{aligned}
\end{equation}
for $x$-direction,
\begin{equation}\label{kekyxy}
	\begin{aligned}
		\alpha_{y}=& \frac{{k}_{y}v_0}{k^2}\left(I_2\otimes \sigma_{x}  {k}_{x}+I_2\otimes \sigma_{y}  {k}_{y}+ \sigma_{x}\otimes I_2  v\Delta k_{x}\right. \\
		&\left.+\sigma_{y}\otimes I_2 v\Delta {k}_{y}\right), \\
		\xi_y=&\frac{\hbar k_x}{2k^2}\left\{I_2\otimes \sigma_{z}\left[\cos(\omega_1 t )-1\right]+ \sigma_{z}\otimes I_2\left[\cos(\omega_2 t)-1\right]\right\}\\
		&+\frac{\hbar k_x}{2k^{3}}\left\{-\left[ I_2\otimes \sigma_{x}k_y-I_2\otimes \sigma_{y}k_x\right]\sin\left(\omega_1 t\right)\right.\\
		&\left.-\left[ \sigma_{x}\otimes I_2 k_y-\sigma_{y}\otimes I_2 k_x\right]\sin\left(\omega_2 t\right)\right\} ,
	\end{aligned}
\end{equation}
for $y$-direction, where $\omega_1=2kv_0 /\hbar$, $\omega_2=2k\Delta \nu v_0 /\hbar$. One can find that $\omega_2$ is 
$\Delta$ times of $\omega_1$. Under the same condition of spinor $\Phi=(0,1,0,1)^T$, the expression of $\bar{\xi}_x$ and $\bar{\xi}_y$ reads
%%%
%%%
\begin{equation}\label{kekyxi}
	\begin{aligned}
\overline{\xi}_x(t)=&\left\langle\Psi_{k}\left|\xi_{x}(t)\right| \Psi_{k}\right\rangle\\
=&\iint\left\{\frac{\hbar k_y}{k^2}\left[\cos(\omega_1t)-1\right]+\frac{\hbar k_y^2}{k^3}\sin(\omega_2t)\right\}\\
&\times\frac{d^2}{\pi} e^{- d^{2}\left[\left(k_{x}-k_{1}\right)^{2}+\left(k_{y}-k_{2}\right)^{2}\right]}dk_xdk_y,\\
\overline{\xi}_y(t)=&\left\langle\Psi_{k}\left|\xi_{y}(t)\right| \Psi_{k}\right\rangle\\
=&\iint\left\{\frac{\hbar k_x}{k^2}\left[1-\cos(\omega_1t)\right]-\frac{\hbar k_xk_y}{k^3}\sin(\omega_2t)\right\}dk_xdk_y\\
&\times\frac{d^2}{\pi} e^{- d^{2}\left[\left(k_{x}-k_{1}\right)^{2}+\left(k_{y}-k_{2}\right)^{2}\right]}dk_xdk_y.\\
	\end{aligned}
\end{equation}
%%%
%%%

Considering the parity of the function, one can find that the integral of an odd power $k_y$ is zero. Then, there is only $\omega_2$ in $x$-direction, and $\omega_1$ in $y$-direction, which agrees well with the numerical results.

Since the matrices in Kek-M Hamiltonian do not satisfy the conditions of Clifford algebra, one cannot get the analytical expression by the same method we used above. Then, for the case of Kek-M, the general approach of ZB is needed to calculate the evolution~\cite{GDavid2010,XShen2022}
%%%
%%%
\begin{equation}
	\begin{aligned}
		\mathbf{A}(t) &=U^{\dagger} \mathbf{A}(0) U=U^{\dagger}[\mathbf{A}(0), U]+U^{\dagger} U \mathbf{A}(0),
	\end{aligned}
\end{equation}
for position operator, there is
\begin{equation}
	\mathbf{{r}}(t)=\mathbf{{r}}(0)+i \hbar U^{\dagger} \frac{\partial U}{\partial \boldsymbol{k}},
\end{equation}
where Hamiltonian can be expressed as $H=\sum_{n}\rho_n E_n$, $E_n$ are the eigenvalues of the Hamiltonian, $\rho_n$ is the density matrix corresponding to $E_n$,  $\rho_m\rho_n=\delta_{mn}\rho_m\rho_n$. According to Eq.~\eqref{coodi}, there are
\begin{equation}\label{orth}
	\begin{aligned}
\boldsymbol{\hat{\alpha}}=\sum_n\rho_n\frac{\partial E_{n}(\boldsymbol{k})}{\partial \boldsymbol{k}},
	\end{aligned}
\end{equation}
\begin{equation}
	\begin{aligned}\label{xM}
		\boldsymbol{\hat{\xi}}=\sum_{m} \sum_{n<m} \boldsymbol{\beta}_{mn}\left[\cos\left(\omega_{mn}t\right)-1\right]+\boldsymbol{\gamma}_{mn}\sin\left(\omega_{mn}t\right),
	\end{aligned}
\end{equation}
where $\omega_{mn}=(E_m-E_n)/\hbar$,
\begin{equation}\label{bgM}
	\begin{aligned}
		\boldsymbol{\beta}_{mn}&=\frac{i \hbar}{E_{m}-E_{n}}\left[\rho_{n} \frac{\partial H(k)}{\partial \boldsymbol{k}} \rho_{m}-\rho_{m} \frac{\partial H(k)}{\partial \boldsymbol{k}} \rho_{n}\right],\\
		\boldsymbol{\gamma}_{mn}&=\frac{\hbar}{E_{m}-E_{n}}\left[\rho_{n} \frac{\partial H(k)}{\partial \boldsymbol{k}} \rho_{m}+\rho_{m} \frac{\partial H(k)}{\partial \boldsymbol{k}} \rho_{n}\right] \text {. }
	\end{aligned}
\end{equation}

For case of Kek-M, the eigenvalues and corresponding density matrix can be written as
%对于Kek-M相，我们写出其本征能量和密度矩阵的表达式：
%%%
%%%
\begin{equation}
	\begin{aligned}\label{eigen}
		E_1=-\Delta  m_0v_0^2-\Omega_2,~E_2=\Delta  m_0v_0^2-\Omega_1,\\
		E_3=-\Delta  m_0v_0^2+\Omega_2,~E_4=\Delta  m_0v_0^2+\Omega_1,
	\end{aligned}
\end{equation}
\begin{equation}\label{density1}
	\begin{aligned}
		\rho_1&=	\frac{1}{4\Omega_2}
		\left(\begin{array}{cccc}
			\Omega_2-z_2 & -v_{0} k_{-} &  v_{0} k_{+} & 		z_2-\Omega_2  \\
			-v_{0} k_{+} & 	\Omega_2+z_2 & 	-\frac{k_+^2v_0^2}{\Omega_2-z_2} &- v_{0} k_{+} \\
			v_{0} k_{-}  & 	-\frac{k_-^2v_0^2}{\Omega_2-z_2} & 	\Omega_2+z_2 & -v_{0} k_{-} \\
			z_2-\Omega_2 & v_{0} k_{-} & -v_{0} k_{+} & 		\Omega_2-z_2
		\end{array}\right),
	\end{aligned}
\end{equation}
%%%
%%%
 \begin{equation}\label{density2}
	\begin{aligned}
		\rho_2&=	\frac{1}{4\Omega_1}
		\left(\begin{array}{cccc}
			\Omega_1-z_1 & -v_{0} k_{-} &  -v_{0} k_{+} & 		\Omega_1-z_1  \\
			-v_{0} k_{+} & 	\Omega_1+z_1 & 	\frac{k_+^2v_0^2}{\Omega_1-z_1} & -v_{0} k_{+} \\
			-v_{0} k_{-}  & 	\frac{k_-^2v_0^2}{\Omega_1-z_1} & 	\Omega_1+z_1 & -v_{0} k_{-} \\
			\Omega_1-z_1 & -v_{0} k_{-} & -v_{0} k_{+} & 		\Omega_1-z_1
		\end{array}\right),
	\end{aligned}
\end{equation}

  \begin{equation}\label{density3}
	\begin{aligned}
		\rho_3&=	\frac{1}{4\Omega_2}
		\left(\begin{array}{cccc}
			z_2+\Omega_2 & v_{0} k_{-} &  -v_{0} k_{+} & 		-z_2-\Omega_2  \\
			v_{0} k_{+} & 	\Omega_2-z_2 & -\frac{k_+^2v_0^2}{z_2+\Omega_2} &- v_{0} k_{+} \\
			-v_{0} k_{-}  &  	-\frac{k_-^2v_0^2}{z_2+\Omega_2} &	z_2-\Omega_2 & v_{0} k_{-} \\
			-z_2-\Omega_2 & -v_{0} k_{-} & v_{0} k_{+} & 		z_2+\Omega_2
		\end{array}\right),
	\end{aligned}
\end{equation}

  \begin{equation}\label{density4}
	\begin{aligned}
		\rho_4&=	\frac{1}{4\Omega_1}
		\left(\begin{array}{cccc}
			z_1+\Omega_1 & v_{0} k_{-} &  v_{0} k_{+} & 		z_1+\Omega_1  \\
			v_{0} k_{+} & 	z_1-\Omega_1 & 	\frac{k_+^2v_0^2}{z_1+\Omega_1} & v_{0} k_{+} \\
			v_{0} k_{-}  & 	\frac{k_-^2v_0^2}{z_1+\Omega_1} & 	z_1-\Omega_1 & v_{0} k_{-} \\
			z_1+\Omega_1 & v_{0} k_{-} & v_{0} k_{+} & 		z_1+\Omega_1
		\end{array}\right),
	\end{aligned}
\end{equation}
where $z_1=m_0v_0^2\left(1+\Delta \right)$,  $z_2=m_0v_0^2\left(1-\Delta \right)$, $\Omega_1=\sqrt{v_0^2k^2+z_1^2}$,  $\Omega_2=\sqrt{v_0^2k^2+z_2^2}$. By plugging Eq.~\eqref{eigen}, \eqref{density1}, \eqref{density2}, \eqref{density3} and \eqref{density4} into Eq.~\eqref{orth}, \eqref{xM} and \eqref{bgM}, one can obtain
%%%
%%%
\begin{equation}
	\begin{aligned}
		\alpha_{x}=\dfrac{v_0^2k_x}{\Omega_1}(\rho_4-\rho_2)+\dfrac{v_0^2k_x}{\Omega_2}(\rho_3-\rho_1),\\
		\alpha_{y}=\dfrac{v_0^2k_y}{\Omega_1}(\rho_4-\rho_2)+\dfrac{v_0^2k_y}{\Omega_2}(\rho_3-\rho_1),
	\end{aligned}
\end{equation}
%%%
and
%%%
\begin{equation}
	\begin{aligned}
		\beta^x_{mn}&=\frac{i \hbar v_0}{E_{m}-E_{n}}\left[\rho_{n}(I_2\otimes\sigma_x) \rho_{m}-\rho_{m} (I_2\otimes\sigma_x) \rho_{n}\right],
    	\end{aligned}
\end{equation}
%%%
\begin{equation}
	\begin{aligned}
		\beta^y_{mn}&=\frac{i \hbar v_0}{E_{m}-E_{n}}\left[\rho_{n} (I_2\otimes\sigma_y) \rho_{m}-\rho_{m} (I_2\otimes\sigma_y) \rho_{n}\right],
  	\end{aligned}
\end{equation}
%%%
\begin{equation}
	\begin{aligned}
		\gamma^x_{mn}&=\frac{\hbar v_0}{E_{m}-E_{n}}\left[\rho_{n} (I_2\otimes\sigma_x)\rho_{m}+\rho_{m} (I_2\otimes\sigma_x) \rho_{n}\right],
  	\end{aligned}
\end{equation}
%%%
\begin{equation}
	\begin{aligned}
		\gamma^y_{mn}&=\frac{\hbar v_0}{E_{m}-E_{n}}\left[\rho_{n} (I_2\otimes\sigma_y)\rho_{m}+\rho_{m} (I_2\otimes\sigma_y) \rho_{n}\right].
	\end{aligned}
\end{equation}
Therefore, in Kek-M system, there are theoretically six frequencies of ZB. By considering the degeneracy that discussed in Appendix~\ref{AP1}, one can find that only four of the six frequencies are observable in experiments. 
%%%
%%%

%\bibliography{kekgr_refer.bib}

\end{document}